\documentclass[journal]{vgtc}                

\usepackage{mathptmx}
\usepackage{mathrsfs}
\usepackage{graphicx}
\usepackage{times}
\usepackage{amsmath}
\usepackage{amssymb}
\usepackage{amsfonts}
\usepackage{algorithm}
\usepackage{subfigure}
\usepackage[american]{babel}
\usepackage{color, colortbl}
\usepackage{paralist}
\usepackage{xurl}
\usepackage{booktabs}
\usepackage{multirow}
\usepackage{changepage}
\usepackage{moresize}
\usepackage[numbers,sort]{natbib}
\usepackage{algorithm}
\usepackage{algorithmicx}
\usepackage{balance}
\usepackage{algpseudocode}
\usepackage{soul}
\usepackage{xspace}
\algrenewcommand\algorithmicrequire{\textbf{Input:}}
\algrenewcommand\algorithmicensure{\textbf{Output:}}

\definecolor{darkred}{RGB}{168,37,17}
\definecolor{darkblue}{RGB}{23,55,119}
\definecolor{highblue}{RGB}{20, 20, 180}
\definecolor{lightblue}{RGB}{183,210,237}
\definecolor{gray}{RGB}{100,100,100}
\definecolor{lightgray}{RGB}{230,230,230}
\definecolor{rainforest}{RGB}{3,101,100}
\definecolor{darkpurple}{RGB}{66,8,91}
\definecolor{orange}{RGB}{242, 101, 34}
\definecolor{black}{RGB}{0, 0, 0}

\newcommand{\new}[1]{#1}

\newcommand{\etal}[1]{et al.~\cite{#1}}

\newcommand{\old}[1]{\xspace}

\onlineid{1139}

\vgtccategory{Research}
\vgtcpapertype{Algorithm/Technique}

\vgtcinsertpkg

\title{Text-to-Viz: Automatic Generation of Infographics from\\ Proportion-Related Natural Language Statements}

\author{Weiwei Cui, Xiaoyu Zhang, Yun Wang, He Huang, Bei Chen,\\Lei Fang, Haidong Zhang, Jian-Guan Lou, and Dongmei Zhang}
\authorfooter{
\item{W. Cui, Y. Wang, H. Huang, B. Chen, L. Fang, H. Zhang, J. Lou, and D. Zhang are with Microsoft Research Asia. Emails: \{weiweicu, wangyun, rayhuang, beichen, leifa, haizhang, jlou, and dongmeiz\}@microsoft.com.}
\item{X. Zhang is with the ViDi Research Group in University of California, Davis, and this work was done during an internship at Microsoft Research Asia. Email: xybzhang@ucdavis.edu.}
}

\shortauthortitle{Biv \MakeLowercase{\textit{et al.}}: Global Illumination for Fun and Profit}

\abstract{
Combining data content with visual embellishments, infographics can effectively deliver messages in an engaging and memorable manner\old{, hence gaining tremendous popularity in many areas, such as business, finance, and health-care}.
Various authoring tools have been proposed to facilitate the creation of infographics. 
However, creating a professional infographic with these authoring tools is still not an easy task, requiring much time and design expertise.
Therefore, these tools are generally not attractive to casual users, who are either unwilling to take time to learn the tools or lacking in proper design expertise to create a professional infographic.
In this paper, we explore an alternative approach: to automatically generate infographics from natural language statements.
We first conducted a preliminary study to explore the design space of infographics. Based on the preliminary study, we built a proof-of-concept system that automatically converts statements about simple proportion-related statistics to a set of infographics with pre-designed styles. Finally, we demonstrated the usability and usefulness of the system through sample results, exhibits, and expert reviews.
}

\keywords{Infographics, automatic visualization.}

\CCScatlist{ 
 \CCScat{K.6.1}{Management of Computing and Information Systems}%
{Project and People Management}{Life Cycle};
 \CCScat{K.7.m}{The Computing Profession}{Miscellaneous}{Ethics}
}

\teaser{
  \centering
  \includegraphics[width = \linewidth]{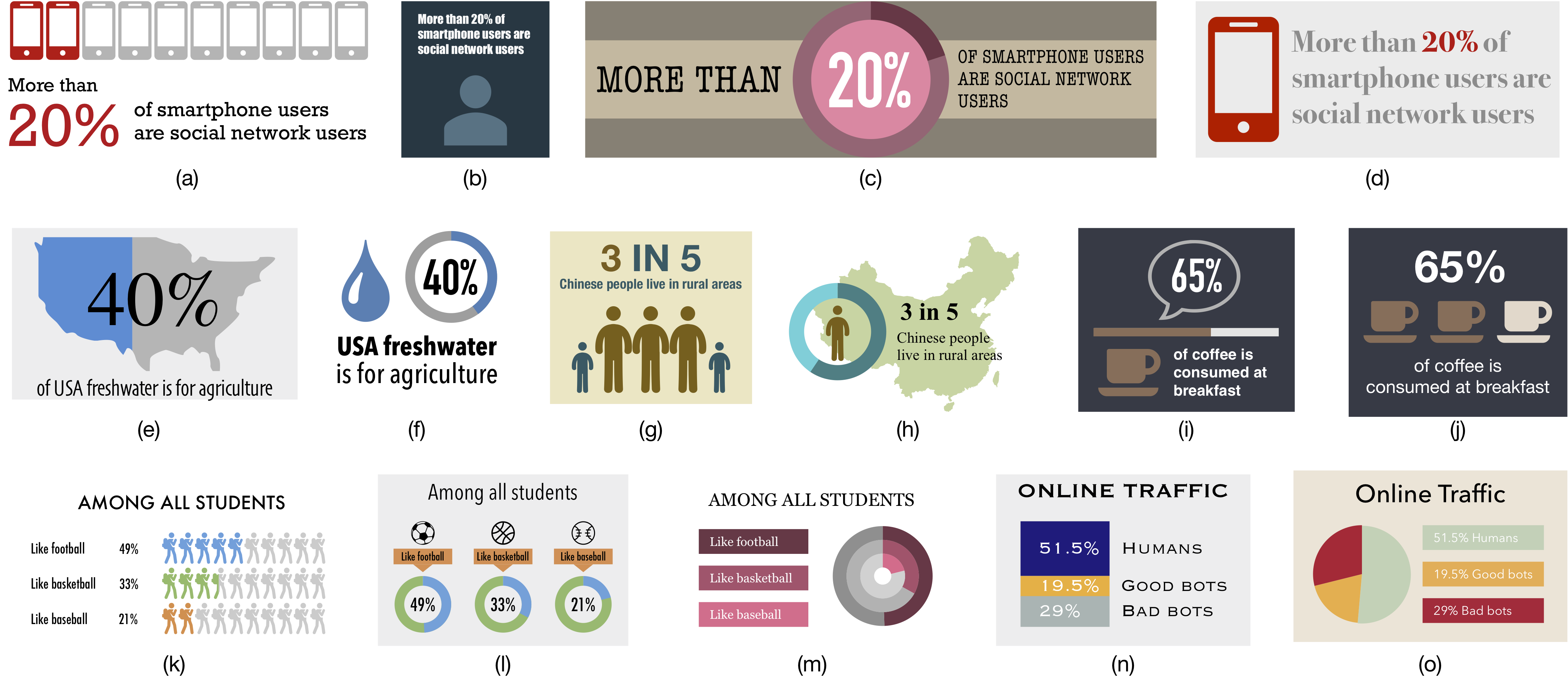}
	\caption{Examples created by Text-to-Viz. (a)-(d) are generated from the statement: ``More than 20\% of smartphone users are social network users.'' (e) and (f) are generated from the statement: ``40 percent of USA freshwater is for agriculture.'' (g) and (h) are generated from the statement: ``3 in 5 Chinese people live in rural areas.'' (i) and (j) are generated from the statement: ``65\% of coffee is consumed at breakfast.'' (k)-(m) are generated from the statement: ``Among all students, 49\% like football, 32\% like basketball, and 21\% like baseball.'' (n) and (o) are generated from the statement: ``Humans made 51.5\% of online traffic, while good bots made 19.5\% and bad bots made 29\%.''}
  \label{fig_examples}
}

\begin{document}
\maketitle

\section{Introduction}
\label{sec_intro}
Information graphics (a.k.a. infographics) is a type of data visualization that combines artistic elements (e.g. clip arts and images) with data-driven content (e.g. bar graphs and pie charts) to deliver information in an engaging and memorable manner~\cite{harrison2015infographic}.
Due to these advantages, they are widely used in many areas, such as business, finance, and health-care, for advertisements and communications.
However, creating a professional infographic is not an easy task.
It is a time-consuming process and also often requires designer skills to ensure the perceptual effectiveness and aesthetics.
 
Much research has been devoted to investigating the design aspect of infographics~\cite{hullman2011impact,skau2017readability,borkin2013makes} and developing authoring tools~\cite{satyanarayan2014lyra,kim2017data,wang2018infonice} to facilitate the creation of data-driven infographics.
Based on different considerations, these authoring tools all strive to reach a balance between the ease-of-use and the power of features, and then to speed up the authoring process.
However, these tools generally target advanced users.
\old{For example, InfoNice~\cite{wang2018infonice} provides a comprehensive graphical user interface to allow manual controls of all visual details. 
Data-Driven Guides~\cite{kim2017data} adopts lazy data-binding to better match the authoring flow of professional designers.}
With complicated editing operations and technical concepts, these tools are not friendly to casual users, who, we believe, form another major category of infographic creators, other than professionals, such as graphic designers and data scientists~\cite{liu2018data}.

Consider a hypothetical example in which a program manager, Nina, is preparing a presentation for her manager and wants to emphasize in her slides that ``40\% of US kids like video games.'' She decides to add an infographic next to the statement with an authoring tool, e.g., DDG~\cite{kim2017data} or InfoNice~\cite{wang2018infonice}.
Since Nina is not a professional graphic designer, she first needs to spend time (re-)familiarizing herself with the tool, such as the user interface and work flow.
However, even if she were familiar with the tool, she still may not know where to begin to create a desired infographic, because all the existing authoring tools assume that the users have a clear or rough idea of what the final infographic may look like.
Unfortunately, Nina has no design expertise and has little to no knowledge of how a professional infographic would look like. 
Therefore, she likely needs to go through existing well-designed infographics (in books or on the Internet) to look for inspiration.
Based on the examined samples, she then settles on a design choice that has the best ``return of investment'' in terms of authoring time and her purpose of emphasizing the message.

From this example, we can summarize some common patterns for this user category.
First, creators in this category only occasionally create infographics and thus are not proficient in the authoring tools.
Second, they do not aim for the most creative/impressive infographics, which often involve complex designs and a long authoring time and are unnecessary in terms of ``return of investment''.
Instead, something effective but professional is often sufficient for their purposes.
Third, they often only have little design expertise, and would likely be unclear on how to design a decent infographic from scratch.
On the other hand, if they were provided with good and relevant samples, they could often quickly pick one based on their personal preferences.

To address the needs of users in this category, we explored a new approach: to automatically generate infographics based on text statements.
Since text is the most common form of communication to exchange information, we believe that this approach can help more people take advantage of infographics.
To achieve this goal, there are two major challenges to overcome.
The first one is to understand and extract appropriate information from a given statement.
The second one is to construct professional infographics based on the extracted information.
For the text understanding challenge, we first collected a corpus of real-world samples.
Then, these samples were manually labeled to train a CRF-based model to identify and extract information for infographic constructions.
For the infographic construction challenge, we analyzed and explored the design space of infographic exemplars that we collected from the Internet.
Based on the design space, we proposed a framework to systematically synthesize infographics.

Considering the numerous types of information that can be represented by infographics~\cite{siricharoen2013infographics,purchase2018classification}, and the numerous ways to express the same information textually and visually, it is impossible to cover the entire space in one paper.
\new{Instead, we decided to focus on a relatively small and isolated text-infographic space and build a proof-of-concept system for it.
To achieve this goal, we first conducted a preliminary survey on the existing infographics to identify a category of information that is commonly represented by infographics and also has clear textual and visual patterns to process systematically.
Based on the preliminary survey, we chose a subtype of information related to proportion facts (e.g., ``Less than 1\% of US men know how to tie a bow tie.'') and built an end-to-end system to automatically convert simple statements containing this type of information to a set of infographics with pre-designed styles.
Finally, we demonstrate the usability and usefulness of the system through sample results, two public exhibits, and expert reviews.}

\old{Instead, we present a proof-of-concept implementation to illustrate this approach for a subtype of information related to proportion facts (Section~\ref{sec:proportion}), e.g., ``Less than 1\% of US men know how to tie a bow tie.''
Finally, we demonstrate the usability and usefulness of the system through sample results, two public exhibits, and expert reviews.}

\section{Related Work}
\old{This section summarizes relevant prior work with regards to information graphics, natural language interactions, auto-generation of data visualizations, and infographic authoring tools.}

\subsection{Information Graphics}
Infographics are a type of data visualization dedicated to presentations and communications, rather than data exploration~\cite{moere2011role,kosara2016presentation}.
By combining data content with visual embellishments, they can effectively deliver complex messages in a memorable and engaging manner~\cite{bateman2010useful,haroz2015isotype,harrison2015infographic}.

Traditional research in infographics mainly focuses on the effects of visual embellishments and understanding their role as an attention-getting device~\cite{skau2017readability,harrison2015infographic,borkin2016}.
Although Tufte~\cite{tufte2001visual} argued that visual embellishments might be harmful, many other studies~\cite{borgo2012empirical,borkin2016,borkin2013makes,moere2012evaluating} have shown that appropriate embellishments, such as pictographs~\cite{haroz2015isotype}, annotations~\cite{ren2017chartaccent}, and imageries~\cite{bateman2010useful,byrne2016acquired}, do not seem to interfere with the correct perception of information, and can increase the long-term memorability of visualizations.
For example, Haroz \etal{haroz2015isotype} demonstrated that embellished charts are equivalent to plain charts in terms of reading speed and accuracy, but the added visual elements make them more memorable.
\old{Bateman \etal{bateman2010useful} also showed evidence that visual embellishments may improve long-term recall without causing noticeable impact on data comprehension.}
Recently, studies have been conducted to directly understand infographics.
For example, Bylinskii \etal{bylinskii2017understanding} adopted OCR techniques to assign hashtags to infographics for information retrieval purposes. 
Madan \etal{madan2018synthetically} used computer vision techniques to detect icons in infographics and proposed a solution to automatically summarize infographics.

\new{Although these studies have extensively demonstrated the value of infographics from various perspectives, and started to investigate infographic understanding in general, none of them shares the same goal of ours, which is to build data-driven infographics automatically for general users.}

\subsection{Natural Language Interactions}
Research on natural language processing~\cite{winograd1972understanding} provides effective ways to analyze texts. 
Syntax analysis~\cite{de2006generating} and semantic analysis~\cite{berant2013semantic} can be used to discover hierarchical structures and understand meanings in text, but they are more appropriate for general language understanding problems. 
Sequence tagging tasks, such as Part-Of-Speech tagging~\cite{marquez1998part} and Named Entity Recognition (NER)~\cite{nadeau2007survey}, aim to assign a categorical label to each word in text sequences, after that, text segments of different types can be obtained based on assigned categorical labels.
Since the input of our system requires extracting text segments, we adopt the sequence tagging techniques.
For sequence tagging techniques, rule-based methods~\cite{rau1991extracting} are straightforward, but costly and limited. 
There are also many machine learning methods proposed which are more robust and easier to generalize, including Hidden Markov Models (HMM)~\cite{bikel1997nymble}, Support Vector Machines (SVM)~\cite{asahara2003japanese} and Conditional Random Fields (CRF)~\cite{lamm2018textual}. 

Recently, human languages have been used as an interface for visual data analysis.
Systems like Articulate~\cite{sun2010articulate}, DataTone~\cite{gao2015datatone}, and Eviza~\cite{setlur2016eviza} adopt various methods to transpile natural language queries to formal database queries, such as SQL, and extract analysis results accordingly.
Although similar to these solutions in terms of understanding user inputs, our system does not relate to database querying.
Instead, we directly visualize user inputs. 
Specifically, we extract essential semantics from user inputs and translate them into infographic representations for visual data storytelling.
In our prototype, we have developed CRF with Convolutional Neural Networks (CNN) because of its high accuracy and efficiency of prediction.

\subsection{Auto-Generated Data Visualizations}
A group of visualization tools enables automatic generation of data visualization through a set of chart templates, including Tableau\footnote{https://public.tableau.com}, Power BI\footnote{https://powerbi.microsoft.com/}, Many Eyes~\cite{viegas2007manyeyes}, and Voyager~\cite{wongsuphasawat2016voyager}. 
Analysts can specify subsets of data to generate visualizations directly.
These tools have eased the exploration of data through quick chart generation. 
Although these tools allow users to easily generate visualizations within a few mouse clicks, users still have to make decisions about what charts to show and how data items are encoded visually.

\new{More recently, machine learning models are leveraged to improve the quality of generated visualizations. 
These solutions may adopt machine learning techniques at different stages of the process. 
Some of them, such as DataSite~\cite{cui2019datasite}, Foresight~\cite{demiralp2017foresight}, and Voder~\cite{srinivasan2018augmenting}, try to automatically extract significant insights from raw datasets, while other ones, such as Show Me~\cite{mackinlay2007show}, DeepEye~\cite{luo2018deepeye}, Draco~\cite{moritz2018formalizing}, and VizML~\cite{hu2018vizml}, use algorithms to identify the best visual form for a given data pattern.
Although all these systems have reduced the effort required for creating data queries and visual encodings, the generated data visualizations are usually standard charts, which are not very expressive and customizable.
In contrast to these systems, we have different focuses at both stages, which is to convert the information that is explicitly provided by users to infographic-style visualizations.}

%
\new{To improve the expressiveness of generated standard charts, some systems also incorporate algorithms to automatically add visual cues~\cite{kong2017internal}.
For example, Contextifier~\cite{hullman2013contextifier} and Temporal Summary Images~\cite{bryan2017temporal} support automatic generation of annotations.
In a broader sense, most visual analytics systems also can automatically generate visualizations (often customized) from their targeting datasets for users to explore and analyze.
However, these tools all require specific types of data format and aim at professional users trying to reveal complex patterns.}

\new{Our system also falls into this category.
However, comparing with these existing tools that aim at standard charts or are deeply coupled with strict data formats, our system has very different input and output, helping users who know exactly what information to visualize and help them convert the information into infographics effortlessly.}

\old{Our approach also falls into this category.
However, unlike existing tools that aim at revealing complex patterns in various datasets, our system aims to help users who know exactly what information to visualize and help them convert the information into infographics effortlessly.
}

\subsection{Infographic Authoring Tools}
Numerous interactive design tools have been proposed to help author more creative and customized infographics. For example, Lyra~\cite{satyanarayan2014lyra}, iVisDesigner~\cite{ren2014ivisdesigner}, and Charticulator~\cite{ren2018charticulator} improve the flexibility of visualization creation through easy manipulation of low-level graphical specifications or declarative models with graphical user interfaces.
iVoLVER~\cite{mendez2016ivolver} and Data Illustrator~\cite{liu2018data} design easy data binding and direct manipulation of graphical widgets to help users automatically transform visual elements to reflect underlying datasets.

To create more customized and engaging infographics, hand-drawn or uploaded images are introduced to improve visualization authoring systems. 
For example, Data-Driven-Guides~\cite{kim2017data} supports vector drawing and data binding, thus providing an authoring environment for data-driven infographics. 
DataInk~\cite{xia2018dataink} enables the creation of infographics through data binding with manipulation of pen and touch interactions.
InfoNice~\cite{wang2018infonice} allows general users to convert unembellished charts into creative visualizations by manipulating visual marks and creating data-driven infographics easily.

Although these systems have greatly reduce the efforts of creating novel data-driven infographic designs they all require users to be familiar with certain non-trivial concepts, such as \textit{vector}, \textit{layer} and \textit{data-binding}, which are often technique barriers for general users to master these systems.
In addition, users need to have an initial idea of the final infographic design and realize the idea step by step, which is also challenging for inexperienced users.
Therefore, these tools, by design, target at professional users, such as designers and data scientists.

\new{In contrast to the existing authoring tools, our system specifically targets users without design experience.
By dramatically reducing the complexity of the authoring process, with the cost of flexibility, our system can help general users create a variety of infographics automatically, saving them time and inspiring their creativity.}

\section{Preliminary Survey of Infographics}
\label{sec_preli}
\begin{figure}
	\includegraphics[width=1\linewidth]{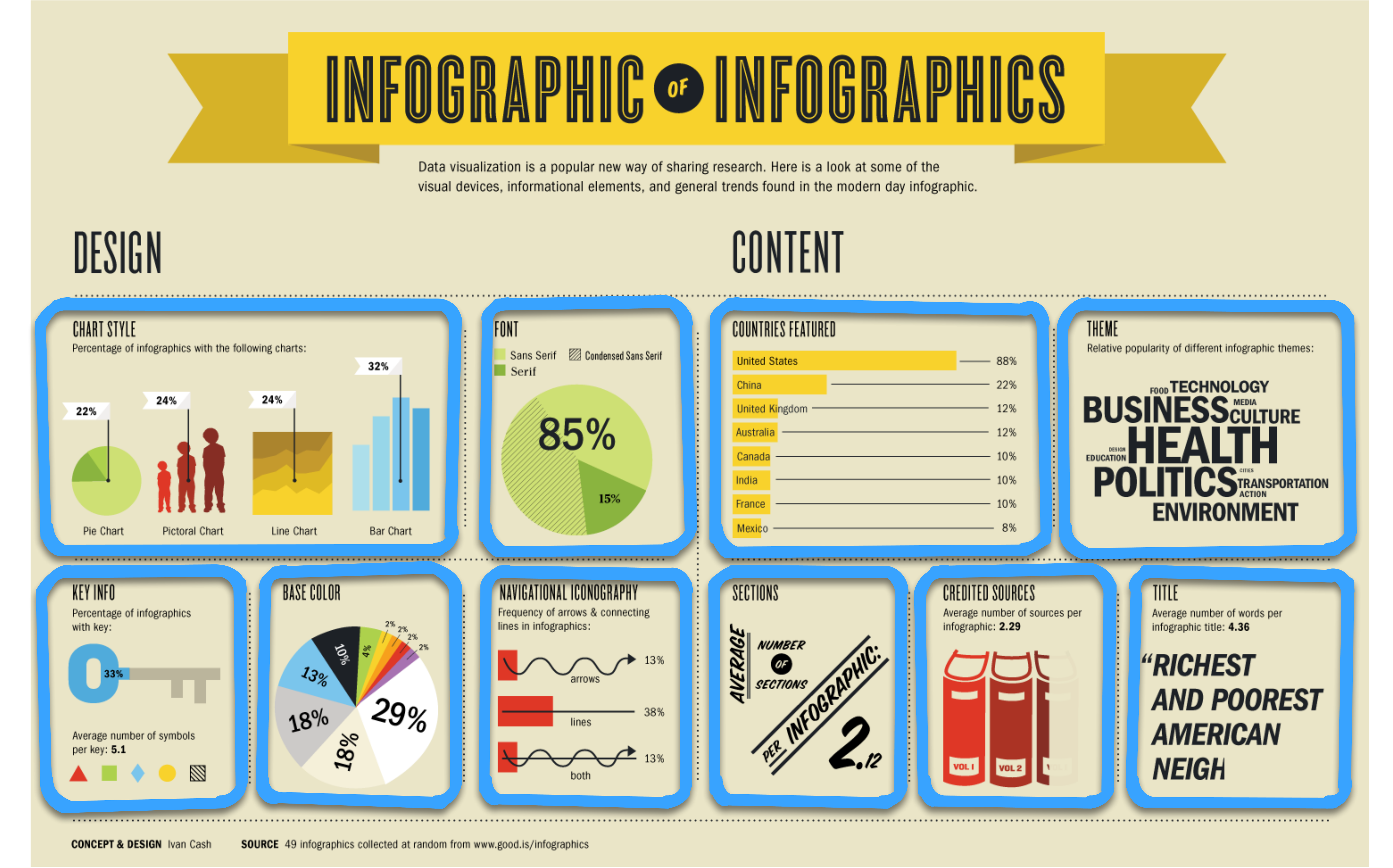}
	\caption{Example of breaking a search result (Infographic of Infographics~\cite{ivancash}) down into individual valid infographic units. All valid units are marked with blue rectangles.}
	\label{fig:infographic_unit}
\end{figure}

The goal of this survey was to better understand how infographics are used in real life and identify a specific type on which to build our proof-of-concept system.
Following the methodology used by Segel and Heer~\cite{segel2010narrative}, we first harvested a representative corpus of infographics from the Internet.
We used ``infographic'' as a search keyword and downloaded the first 200 distinct results manually from Google Image Search.
Due to Google's ranking mechanism, the top query results were believed to have good quality and relevance.
Since in each search result, multiple related infographics may be arranged together for a comprehensive story, we further manually broke them down into 983 individual infographic units (Figure~\ref{fig:infographic_unit}) based on the following criteria. A valid infographic unit should:
\begin{compactitem}
	\item deliver at least one message;
	\item contain at least one graphical element;
	\item be visually and semantically coherent and complete;
	\item not be able to split into smaller units that  fulfill the above criteria.
\end{compactitem}

\begin{table}[h!]
	\begin{center}
		\begin{tabular}{c|c|r}
			\toprule
			\multicolumn{2}{c|}{Category} & Percentage\\
			\midrule
			\multirow{6}{*}{\begin{tabular}[c]{@{}c@{}}Statistics-based\end{tabular}} & Proportion & 45.7\%\\
			{} & Quantity & 28.3\%\\
			{} & Change   & 8.5\%\\
			{} & Rank     & 5.8\%\\
			{} & Other    & 6.2\%\\
			\midrule
			\multicolumn{2}{c|}{Timeline-based} & 1.5\%\\					
			\midrule
			\multicolumn{2}{c|}{Process-based} &  1.1\%\\
			\midrule			
			\multicolumn{2}{c|}{Location-based} & 0.7\%\\
			\midrule			
			\multicolumn{2}{c|}{Other} & 2.2\%\\
			\bottomrule		
		\end{tabular}
	\end{center}
	\caption{Information categories of the collected infographics. Please note that the percentages are based on 983 infographic units, rather than 200 infographic sheets. Therefore, the numbers for non-statistics based infographics are exceptionally low, since they often take a larger space than statistics-based ones. \old{In other words, the number 983 is diluted with statistics-based units, since we often extract much more of them than non-statistics ones from a sheet of a typical size.}}
	\label{tab:table1}
\end{table}

We then categorized all the resulting infographic units into four main groups (Table~\ref{tab:table1}) proposed by Artacho-Ram \etal{artacho2008influence}:
\begin{compactitem}
	\item \textbf{Statistical-based}: This is the main category of infographics, which commonly includes horizontal bar charts, vertical column charts, pictographs, and pie/donut charts, for summarizing a large amount of statistical information.
	\item \textbf{Timeline-based}: This category aims to show information and events happening over time, helping audiences realize chronological relationships quickly. Common visual representations include time-lines, tables, etc.
	\item \textbf{Process-based}: This category aims to instruct readers to take step-by-step actions to achieve a certain goal. They are often used to illustrate cooking recipes in magazines, or to clarify operations in workspaces or factories.
	\item \textbf{Location-based}: Infographics in this category generally contain a map and other annotations such as icons, diagrams, arrows, and tables for navigation. They are usually designed for places such as tourist spots, malls, factories, etc.
\end{compactitem}

Please note that this categorization and the others discussed in this paper were first independently performed by three of the co-authors, and then documented as discussion results among the co-authors.

\begin{figure}
	\includegraphics[width=1\linewidth]{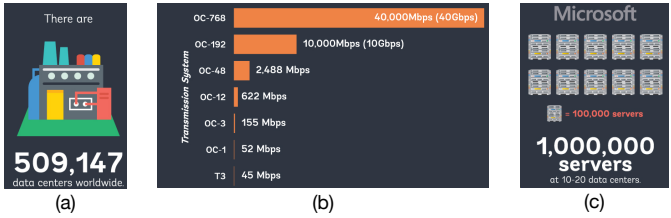}
	\caption{Exemplars of quantity-related infographics~\cite{example_quantity}: (a) embellishment icons, (b) horizontal bar charts, and (c) pictographs.}
	\label{fig_quantity_eg}
\end{figure}

\begin{figure}
	\includegraphics[width=1\linewidth]{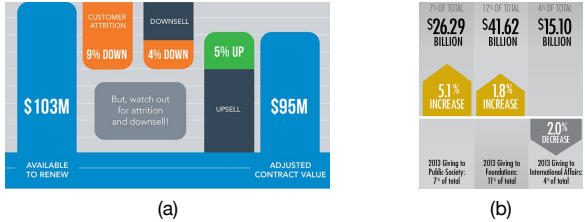}
	\caption{Exemplars of change-related infographics~\cite{example_change2,example_change1}: (a) contrast color + side-by-side comparison and (b) contrast color + arrows.}
	\label{fig_change_eg}
\end{figure}

\begin{figure}
	\includegraphics[width=1\linewidth]{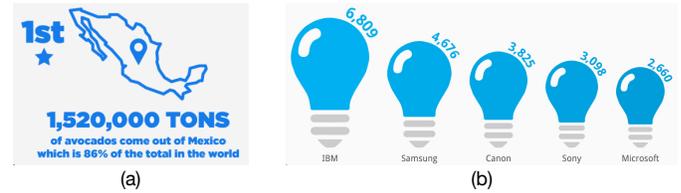}
	\caption{Exemplars of rank-related infographics~\cite{example_rank2,example_rank1}: (a) highlighted keyword + star embellishment and (b) ordered small-multiples.}
	\label{fig_rank_eg}
\end{figure}

When examining infographics in the dominant statistics-based category, we further discovered diverse patterns in terms of graphic designs and the underlying messages, which were categorized into four major sub-categories, namely \textit{proportion}, \textit{quantity}, \textit{change}, and \textit{rank}:

\begin{compactitem}
	\item \textbf{Proportion}: This category of infographics aims to convey statistical information about how much a part occupies the whole, e.g., ``More than \textit{1/3} of the U.S. adult population is obese''.
	According to our survey, this is the most common type of information delivered in the data collected.
	Infographics in this category often have conspicuous visual and textual patterns. 
	Visually, they often contain bar charts, donut charts, pie charts, pictographs, etc.
	Textually, the key information is often expressed in forms like ``\textit{n}\%'', ``\textit{m} in \textit{n} '',  ``\textit{m}/\textit{n}'', or ``half of \ldots''.
	Because of its prevalence in practice and relatively prominent visual and textual features, we decided to build the proof-of-concept system for this infographic category.
	
	\item \textbf{Quantity}:
	The second largest category is related to quantities. 
	These infographics emphasize the quantitative amount of a certain unit, such as income, population, or speed.
	Since the values cannot be described as fractions, visuals such as donut charts and pie charts do not apply here.
	Popular charts in this category include pictographs, vertical column charts, and horizontal bar charts (Figure~\ref{fig_quantity_eg}(b) and (c)).
	However, we also noticed that more samples in this category do not incorporate any data-driven visuals.
	Instead, these infographics mainly contain embellishment visuals (e.g., icons) along with the quantity values highlighted with color, font size/weight, etc. (Figure~\ref{fig_quantity_eg}(a)).

	\item \textbf{Change}:
	The third category aims at describing changes.
	Although values in this category are often also expressed as proportions or quantities, this category is different from the previous two by emphasizing the concept of change, such as ``increase'', ``decrease'', or ``drop''.
	In the corresponding infographics, arrows (Figure~\ref{fig_change_eg}(b)), contrasting colors (Figure~\ref{fig_change_eg}(a) and (b)), and side-by-side comparison (Figure~\ref{fig_change_eg}(a)) are often employed in this category to depict the directions of changes.
	
	\item \textbf{Rank}:
	This category shows the relative position of a data item in a group, which is easy to identify based on ordinal numbers or specific symbols, such as ``\#'', ``No.'' or ``Top $n$''.
	However, the visualizations for single or multiple ranks are different.
	For a message that only contains one piece of ranking information, such as ``Florida has the \textbf{3rd} largest homeless population,'' infographics usually highlight the key words, such as ``\textit{3rd}'', and sometimes incorporate metaphoric embellishments, such as stars, medals, or trophy cups (Figure~\ref{fig_rank_eg}(a)).
	On the other hand, for a message that involves the ranking of multiple data items, infographics often arrange representing icons in an ordered list to directly convey their relative ranks (Figure~\ref{fig_rank_eg}(b)).
\end{compactitem}

In addition to the information categories, we also discovered that the number of statistical facts covered in one infographic unit may vary. 
In many cases, an infographic unit only covers one fact.
But there is also a considerable amount of infographics that contain multiple facts.
Therefore, based on the number of facts and the relationships between facts in an infographic, we also identified four different categories, namely, \textit{single}, \textit{composition}, \textit{comparison}, and  \textit{accumulation}, accounting for 32\%, 25\%, 27\% and 16\% of the collected infographic units respectively.

\begin{compactitem}
	\item \textbf{Single}: There is only one statistical fact in an infographic, it describes one facet of a subject.
	This is the simplest but the most commonly found type in our dataset.
	Since there is only one fact, the corresponding visualization is generally decided by its semantic type (Table~\ref{tab:table1}).
	\item \textbf{Composition}: In this category, an infographic depicts more than one facet of a subject to form a complete picture of it. 
	For example, in the statement, ``In the United States alone, there were 10.5 billion searches in July 2009, which is a 106\% increase from 5.1 billion in Dec. 2005,'' all of the numbers are used to provide information about searches in the United States. 
	Since this category may involve different types of statistical facts, we do not cover it in our prototype system.
	\item \textbf{Comparison}: For this category, multiple facts are provided to compare the same facet of different subjects. 
	Taking the statement ``49\% of students like football, while 33\% of students like basketball'' as an example, numbers ``49\%'' and ``33\%'' compare the students who like football or basketball.
	The facets of these two numbers are the same, i.e., students.
	Therefore, they are either combined into a bar chart or placed side-by-side using the same type of visual element. 
	When visual elements are placed side-by-side, distinguishing colors, sizes, or icons are usually used to show difference and assist in comparison.
	\item \textbf{Accumulation}: Similar to \textit{comparison}, this type of facts also depicts one single facet for different subjects. 
	But these facts can be logically combined to form a larger whole. 
	An example for \textit{accumulation} is this statement, ``60\% of participants come from the US, while 40\% come from Canada.''
	In many cases, the numbers add up to one.
	And because of this characteristic, designers prefer to combine all the data into one pie chart, donut chart, or stacked bar chart, instead of visualizing individual facts separately.	
\end{compactitem}

Please note that the visualizations may overlap for categories \textit{comparison} and \textit{accumulation}. 
For some messages in the category of \textit{accumulation}, designers may still use visualizations for \textit{comparison}, e.g., side-by-side visuals, to emphasize their difference. 
Therefore, the visualizations suitable for \textit{comparison} are also suitable for \textit{accumulation}. 
However, the same is not true vice versa.

\section{Proportion-Related Information}
\label{sec:proportion}
According to our preliminary survey, we decided to target proportion-related facts because of their dominance in our harvested dataset and distinct visual and textual patterns.
In this section, we further analyze the textual and visual spaces of this type of information, on which we can build a model to automatically convert natural language statements into proper infographics.

\subsection{Text Space}
One goal of our system is to hide all technical concepts/details, so that the learning curve is minimized and everyone can easily create infographics without obstacles.
Therefore, it is critical to allow users to provide the information to visualize using their daily language, rather than formal declarative languages, such as JSON or XML.
However, since a message may have many different ways to express in the natural language, we need to collect real-world samples to understand how people usually deliver proportion facts in text.

\new{From a search engine index, we collected a dataset of real-world PowerPoint files that contains approximately 100,000 slides.}
From the slides, we used regular expressions to capture 5,562 statements that contain common expressions for fraction, such as \textit{n\%}, \textit{m} in \textit{n}, \textit{m} out of \textit{n}, and \textit{half of}.
Please note that not all the captured statements can be classified as \textit{proportions} based on the above definitions, since statements such as ``Company revenue increased by 24\%'' are also included here, which, obviously, should belong to \textit{change}.
Therefore, from the collected statements, we further manually sampled 800 valid proportion-related ones, and used them as the training dataset to build a machine learning model for text processing (Section~\ref{text_process}).

At first, we also collected samples from online news articles.
However, comparing statements from these two sources, we found that statements in PowerPoint slides are often shorter and more concise than those in news articles.
Since PowerPoint-style statements are more focused on statistical facts and better match descriptions in infographics, we eventually removed samples from news articles and only kept the ones collected from PowerPoint slides.

\subsection{Visual Space}
To understand the infographics of proportion facts, we held a one-hour discussion with two experts, both of whom have at least four years of experience as freelance graphic designers.
During the interview, we presented examples of proportion-related infographics, and tried to understand how these infographics were generated from scratch based on their experiences.
The discussion resulted in a design space with four dimensions: \textit{layout}, \textit{description}, \textit{graphic}, and \textit{color}.
Although there are a wide variety of designs, every infographic can be described as a tuple of these values. 
And the design space also aligns with the designers' mental model to create infographics.
By thoroughly examining the collected samples, we summarized each dimension as follows.

\subsubsection{Layout}
Based on the existing layout taxonomies on general visualization design~\cite{segel2010narrative,bach2018design}, we analyzed and discussed the layout of infographics for proportional facts.

\paragraph{Layout for Single Fact}
Infographics with single proportion facts are the simplest and most common type based on our preliminary survey.
They are generally composed of two main visual elements: descriptions and graphics. 

\begin{figure}
	\includegraphics[width=1\linewidth]{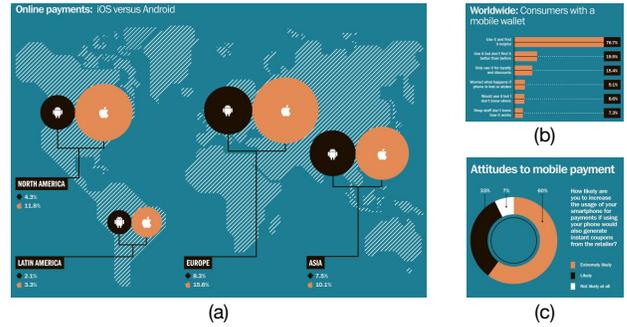}
	\caption{Exemplars of infographics with multiple facts~\cite{example_multiple2}: (a) side-by-side and sharing context, (b)  sharing axes, (c) sharing center.}
	\label{fig_multiple_layout}
\end{figure}

In most cases, the descriptions and graphics are arranged in a grid layout. 
They can be aligned in a horizontal (Figure~\ref{fig_examples}(c) and (d)), vertical (Figure~\ref{fig_examples}(b), (g) and (j)), or tiled (Figure~\ref{fig_examples}(a), (f), and (i)) fashion. 
Another common method is overlaying (Figure~\ref{fig_examples}(e), (h)). For example, when proportion facts contain geographic information, descriptions are usually overlaid on top of a map. 
\paragraph{Layout for Multiple Facts}
When combining multiple facts into one, there are four common strategies:

\begin{compactitem}
	\item \textbf{Side-by-Side}. To indicate comparison, hierarchy, or other logical relationships, multiple infographics can be arranged in grids, including ones that are parallel, tiled, circular, hierarchical, and stacked (Figure~\ref{fig_multiple_layout}(a)).
	\item \textbf{Sharing axes}.  This type of layout aligns the visual marks of all single infographics so that they can be placed in a common coordinate system. It is applicable for statistical charts with axes, such as bar charts or scatterplots (Figure~\ref{fig_multiple_layout}(b)). 
	\item \textbf{Sharing center}. This type of layout arranges multiple numerical facts in the form of concentric circles or sectors with the same center. It is applicable for circular charts, such as pie charts or Nightingale rose charts (Figure~\ref{fig_multiple_layout}(c)). 
	\item \textbf{Sharing context}. This type of layout links multiple annotations to different positions of shared illustrative images. It is applicable to annotated infographics. Visual elements of multiple infographics are placed according to the empty space of a common background (Figure~\ref{fig_multiple_layout}(a)).
\end{compactitem}

\subsubsection{Description}
\label{sec_entity}
Text description is another key component in data-driven infographics. 
The original statements provide basic materials of infographic descriptions. 
Infographic designers usually extract key information and arrange descriptions as text blocks. 
Each infographic may contain multiple text blocks, combined to deliver a complete message.
As mentioned in Section~\ref{sec_preli}, proportion-related facts are about how much a part occupies the whole.
Therefore, there are three key pieces of information for each proportion fact: \textit{number}, \textit{part}, and \textit{whole}, which are often explicitly expressed in infographics as descriptions.
In addition to these three pieces, we also discovered that designers often treat a number's \textit{modifier} (if there is one), such as ``more than'' and ``less than'', as an isolated description.
Besides these four key types of descriptions, we also identify the following common forms:

\begin{compactitem}
	\item \new{The entire statement.}
	\item \new{The statement with the \textit{number} removed}: For example, ``of USA fresh water is used for agriculture'' in the statement ``40 percent of USA fresh water is used for agriculture.''
	\item \new{The \textit{part} as a verb-object phrase in the statement}: Syntactically, when a statement contains a subject and a verb-object phrase, this type of description directly shows the phrase part, for example, ``are consumed in breakfast'' in the statement ``65\% of coffee are consumed in breakfast.''
	\item \new{The \textit{number}-\textit{whole} phrase in the statement}: Basically, this type of description shows the subject in a statement, where the \textit{modifier} is optional, for example, ``65\% of coffee'' in ``65\% of coffee are consumed in breakfast.''
	\item \new{The text segments before and after the \textit{number} in the statement}: Based on the position of the \textit{number} component, a statement can also be segmented into three parts and arranged separately, for example, ``In the US, less than'', ``1\%'', and ``men know how to tie a bow tie'' in ``In the US, less than 1\% men know how to tie a bow tie.'' This segmentation is particularly useful when a statement does not start with the \textit{number} component.
\end{compactitem}

\subsubsection{Graphic}
The design of graphic components involves the selection and composition of graphics.
There are two main considerations for how to display graphics in infographics.
First, graphics should be semantically related to the content of the original statements.
Second, different graphic elements may have different roles in helping to convey the message.

Based on some existing visualization taxonomies~\cite{borkin2013makes, amini2017authoring}, we went through our infographic dataset for proportional facts. We summarized frequently-used visualizations into seven types, namely, pictograph (Figure~\ref{fig_examples}(a)), adornment (Figure~\ref{fig_examples}(d)), donut chart (Figure~\ref{fig_examples}(c)), pie chart (Figure~\ref{fig_examples} (o)), bar chart (Figure~\ref{fig_examples}(i)), filled icon (Figure~\ref{fig_examples}(e)), and scaled icon (Figure\ref{fig_examples}(g)).

\subsubsection{Color}
Each infographic has a color theme consisting of harmonious colors, which may also help indicate latent semantics.
For example, when an infographic is about environment protection, green or blue based themes are often used.
In addition, given a color palette, colors also need to be assigned to different parts of an infographic.
The designs in our samples share common rules:

\begin{compactitem}
	\item \textbf{Background and foreground}: The background color occupies the biggest area and contrasts the foreground color, while there are often multiple foreground colors used for different visual elements, such as the descriptions, title, and icons. 
	In most cases, descriptions in different text blocks share the same color.
	Graphic elements, on the other hand, may use one or more colors based on their roles. For example, pictographs, pie charts, or donut charts often require at least two colors to correctly convey the proportion value. Embellishing icons often only use one color to keep a clean look.
	\item \textbf{Number highlight}:
	Being an important component, \textit{numbers} often require extra emphasis. 
	Most of the samples highlight the values with  a distinguished size, color, and font (e.g., Figure~\ref{fig_examples}(a) and (g)). Alternatively, values can be embellished with background pictures (e.g., Figure~\ref{fig_examples}(c) and (i)). 
\end{compactitem}

\section{Text-to-Viz Implementation}
Our system contains two main modules, namely \textit{text analyzer} and \textit{visual generator}.
First, users provide a textual statement about a proportion fact, such as ``More than 40\% of students like football.''
Then, our \textit{text analyzer} identifies the essential segments in the statement, including \textit{modifier}, \textit{whole}, \textit{part}, \textit{number}, and others (Section~\ref{sec_entity}).
Then the original statement and the extracted segments are fed into the \textit{visual generator} for infographic construction.
For each dimension (i.e., \textit{layout}, \textit{description}, \textit{graphic}, and \textit{color}), a set of visual elements are generated or selected. 
Then, we enumerate all combinations of these elements, to synthesize valid infographic candidates.
Finally, all the synthesized results are evaluated and ranked, and the ones with high scores are recommended to users.
Then, users can either directly export anyone of them as an image to integrate into their reports or presentation slides, or select one and further refine it based on their personal preferences.

\subsection{Text Analyzer}
\label{text_process}
\new{Given a statement on proportion facts, the goal of this analyzer is to extract the segments of four predefined entity types: modifier (M), number (N), part (P), and whole (W), which is a fundamental task in natural language understanding called named entity recognition. Currently CRF-based approaches have the state-of-the-art performance~\cite{ju2018neural}. In this prototype, for efficiency, we develop a supervised CNN+CRF model to perform the task. Specifically, there are three steps:}

\begin{figure}
	\includegraphics[width=1\linewidth]{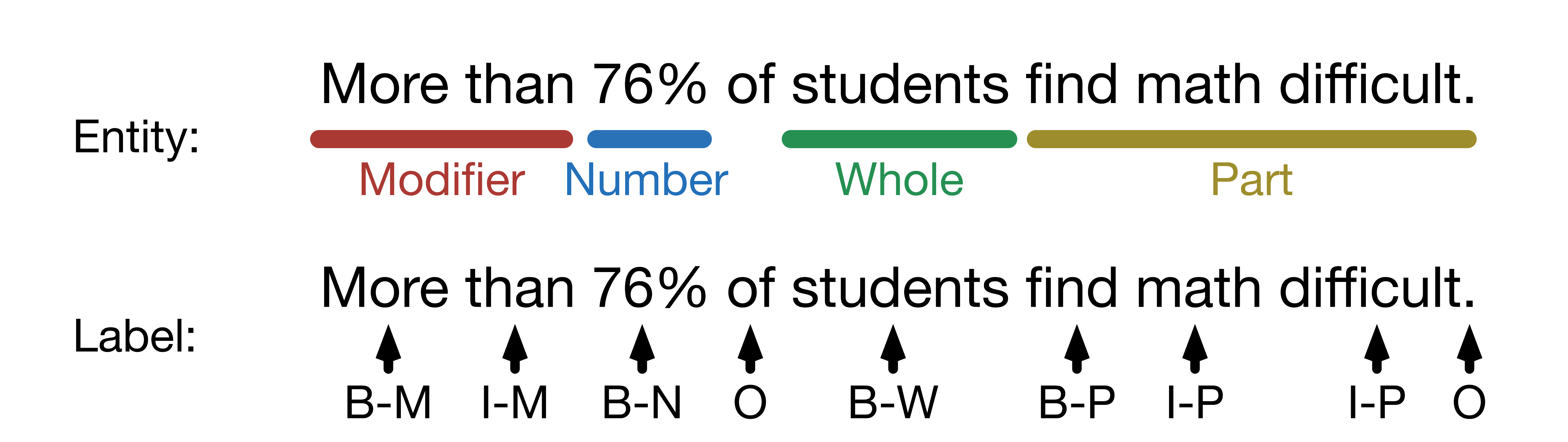}
	\caption{An example of entities and labels in a statement. Following the IOB (inside, outside, and beginning) format~\cite{ramshaw1999text}, we can map the text and entities into a sequence of labels, \new{where B-, I-, and O represent begin, inside, and outside, respectively.}}
	\label{fig:image_ner}
\end{figure}

\old{We formulate our \textit{text analyzer} as a sequence tagging problem with four predefined entity types:
\textit{modifier} (M), \textit{number} (N), \textit{part} (P), and \textit{whole} (W). 
Given a statement on proportion facts, the goal is to extract the segments of these entities. 
An example is shown in Figure~\ref{fig:image_ner}. 
Following the IOB (inside, outside, and beginning) format~\cite{ramshaw1999text}, we map the text and entities into a word label sequence. 
Then a supervised CNN+CRF model is developed to perform word sequences labeling. 
Specifically, there are three steps: 
}

\begin{compactitem}
	\item \textbf{Tokenization}: \new{First, a sentence is converted into a token sequence through preprocessing, where we take punctuations, abbreviations, and special phrases into consideration. For example, given the statement in Figure~\ref{fig:image_ner}, we can totally collect a sequence of nine tokens (i.e.,``more'', ``than'', \ldots, and the final period ``.'').}
	\item \textbf{Featurization}: 
	\new{Then, for each token, we extract multiple features, such as word embedding feature~\cite{mikolov2013efficient}, syntactic feature (such as upper/lower case, punctuation, part-of-speech tag), and Brown clustering feature~\cite{brown1992class}. All these features are concatenated together to form a larger feature vector to represent the token. For the aforementioned example of nine tokens, a $9\times n$ feature matrix can be obtained, where $n$ ($n$=2531 in our implementation) is the length of the concatenated feature vector.}
	\item \textbf{CNN + CRF}: \new{In this step, we feed the output feature matrix to a one-dimensional Convolutional Neural Network (CNN)~\cite{hu2014convolutional}, on top of which is a Conditional Random Field (CRF)~\cite{mccallum2003early} layer. In this step, the CNN layer is used to learn a better feature representation for each token from the input raw feature matrix. For the aforementioned example of nine tokens, our CNN layer will generate a $9\times m$ feature matrix, where $m$ ($m$=59 in our implementation) is the number of kernels in the layer. Then the CRF layer is used to determine the entity labels for all tokens. The parameters of the CNN and CRF are jointly trained.}
	
\end{compactitem}

We train the CNN+CRF model on 800 manually annotated statements.
Table~\ref{tab_cnn_result} shows the performance of 10-fold cross-validation.

\begin{table}
	\begin{center}
			\begin{tabular}{c|cccc}
			\toprule
			Entity & Modifier & Whole & Part & Number\\
			\midrule
			Precision & 1.00 & 0.81 & 0.65 & 1.00\\
			Recall & 0.94 & 0.74 & 0.73 & 0.97\\
			F1 Score & 0.97 & 0.77 & 0.69 & 0.97\\
			\bottomrule
		\end{tabular}
	\end{center}
	\caption{Training results for \textit{text analyzer} model.}
	\label{tab_cnn_result}
\end{table}

\subsection{Visual Generator}
Taking the information provided by the \textit{text analyzer}, this generator identifies multiple candidates on each design space dimension, including  \textit{layout}, \textit{description}, \textit{graphic}, and \textit{color}.
After that, all the combinations of these candidates are enumerated to identify valid infographic results in the \textit{synthesis} step.
Then, the quality of the resulting infographics is evaluated through the \textit{ranking} step.

\subsubsection{Layout}

The \textit{layout} module contains a set of blueprints that describe the overall look of the resulting infographics.
First, a blueprint specifies the aspect ratio of the resulting infographic and how the space is divided into regions. 
More importantly, for each region, the blueprint also needs to specify what kind of information should be placed there, along with any constraints concerning the corresponding information.

\begin{figure}
	\includegraphics[width=1\linewidth]{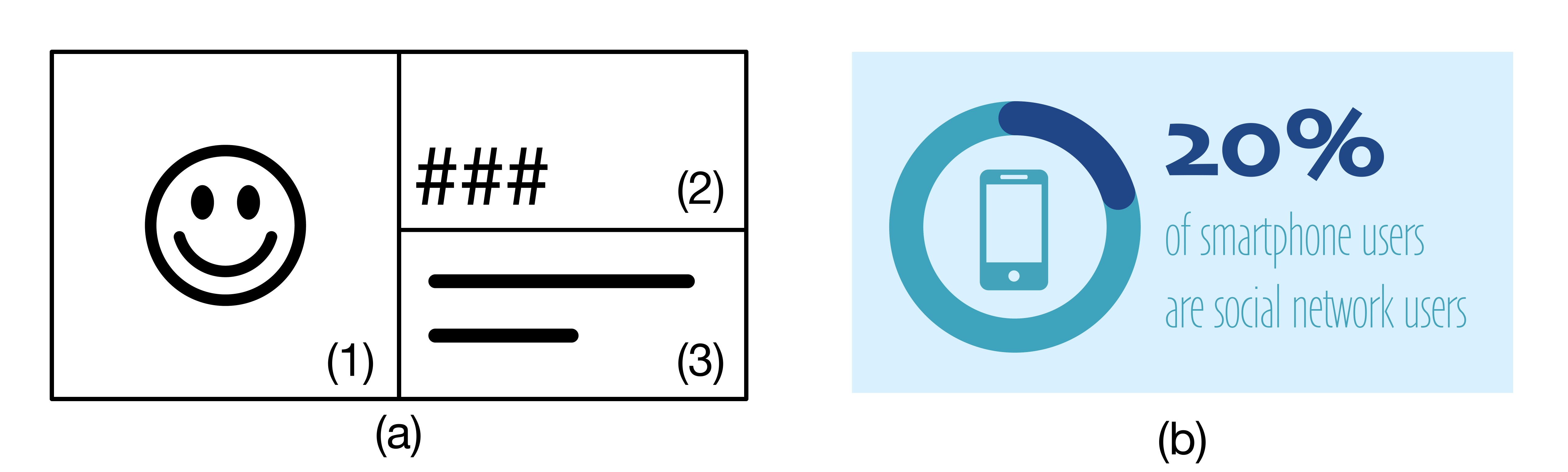}
	\caption{(a) A layout blueprint example and (b) its realization.}
	\vspace{-1mm}
	\label{fig_layout_example}
\end{figure}

Figure~\ref{fig_layout_example}(a) shows an example. 
This blueprint first specifies that the resulting infographic should have an overall $2\times1$ aspect ratio.
The canvas is then recursively divided into three regions.
Although topological relationships are specified, individual regions are allowed to grow or shrink to accommodate their content.
Furthermore, we added the following key constraints:
\begin{compactitem}
	\item This blueprint only accepts statements that start with numerical values, e.g., ``76\% students find math difficult.''
	\item Region 1 holds a graphical element, which can be a pictograph, a filled icon, a donut chart, or a pie chart;
	\item Region 2 holds the \textit{number} part in the input statement;
	\item Region 3 holds the the input with the \textit{number} part removed;
	\item The font size in Region 2 should be at least three times and at most eight times the font size in Region 3.
\end{compactitem}
Clearly, this \textit{layout} module cannot enumerate all creative infographic designs.
However, the goal of this system is to provide the quickest way to generate common but professional infographics.
Thus, we analyzed the collected samples, identified a set of exemplars, and built 20 layout blueprints in total.
These layouts serve as our initial pool of blueprints for infographics generation.
However, this is an expandable approach, so we can add more layouts as needed.

\subsubsection{Description}

The basic descriptions, i.e., \textit{modifier}, \textit{part}, \textit{whole}, and \textit{number}, are identified by the \textit{text analyzer}. 
To provide additional description candidates (Section~\ref{sec_entity}) that may be required by different blueprints, we adopted the Stanford Parser~\cite{de2006generating} to analyze the grammatical structure of the input natural language statement.
From the generated grammar tree, we extracted descriptions with different lengths and different components to meet the need of different layouts

\subsubsection{Graphic}
This \textit{graphic} module aims to provide all graphic elements, including representative icons and other data-driven graphics, such as pie charts and donut charts.
In addition, various applicable constraints should also be considered here.
When integrating icons into layout blueprints, these constraints will decide whether the resulting infographic is a valid one or not.
For example, country-shape icons are not suitable for pictographs and are often used as backgrounds (Figure~\ref{fig_examples}(h)) or filled shapes (Figure~\ref{fig_examples}(e)).
Icons that represent the \textit{part} are not suitable for pictographs either.
Hollow icons also cannot be used as filled shapes, since the colors in them are hardly discernible.
In our implementation, we built an icon library with 100 common icons.
For each icon, we manually added one or more keywords indicating its semantic meaning, and proper applicable constraints.
Then, for non-stop words in the \textit{part} and \textit{whole} components, we use the Word2Vec~\cite{mikolov2013efficient} technique to find the best-matching icons in the icon library.


\subsubsection{Color}
The \textit{color} module aims to find a set of color palettes for a specific infographic.
There are two considerations here. 
First and foremost, the colors within a palette should be harmonic.
In addition, it would be better if the colors match the context of the given statement.
For example, if the proportion fact is about environment protection, it is more appropriate for the color theme of the corresponding infographic to contain green or blue colors.
Although several techniques~\cite{setlur2016linguistic,lin2013selecting} have been proposed to identify semantically relevant colors from words, they mainly focus on individual colors and cannot generate harmonic color palettes.
Therefore, to address the needs of our \textit{color} module, we adopted a different approach and built a theme library, which contains a set of color palettes.
In each palette, various colors are defined with annotations describing their specific uses, such as background, highlights, text, etc.
Leveraging the color design knowledge~\cite{adobecolor,coolor18}, we can ensure that the colors within a palette are harmonic.
In addition, we also added one or more keywords to each color palette, describing their preferred context, such as ``environment'', ``fruit'', ``taxi'', and ``technology''.
Then for the non-stop words in the input statements, we again use the Word2Vec~\cite{mikolov2013efficient} technique to find the best-matching palette.


\subsubsection{Synthesis}
At the \textit{synthesis} step, for each layout blueprint, we enumerate through all recommended graphics, color palettes, and descriptions, and then generate all valid infographics.

First, we rule out all the layout blueprints that require elements non-existent in the input statement. 
For example, if a blueprint requires a region to be filled with \textit{modifier} information (e.g., ``More than''), but such information is not provided by the \textit{description} module, then this blueprint will be considered invalid and ruled out.
Then, for each valid blueprint, we extract the required icon and description information, calculate their aspect rations, and try to put them into the blueprint layout.
The goal is to scale them uniformly and maximize their total size, given the predefined layout constraint.
We formulate it as a UI interface optimization problem and solve it using the solver proposed by Badros \etal{badros2001cassowary}.

Since the final results are optimized based on the aspect ratios of visual elements, we also need to pre-compute several options for pictographs and descriptions. 
For example, we choose several common pictograph arrangement, such as 2$\times$5, 1$\times$5, and 1$\times$10, to obtain different aspect ratios for the solver to enumerate.
Given a description, different line breakings may also yield different aspect ratios.
To obtain a set of candidate aspect ratios for a description, we first consider the different line counts (from 1 to 10).
Then, we use dynamic programming to obtain a line breaking setting that has the minimum raggedness~\cite{knuth1981breaking} for each line count, since designers generally prefer similar lengths when a text is broken into multiple lines.
\new{In addition, a pre-selected set of fonts with different compactnesses are also enumerated to find the font that yields the best aspect ratio to match the assigned canvas space.}

\subsubsection{Ranking}

Since there may be multiple values recommended by these modules, the number of valid combinations may be large. And clearly they are not equally appealing.
Therefore, for each resulting infographic, we need to evaluate its quality of message delivery.
In this system, we consider three scores, namely \textit{semantic}, \textit{visual}, and \textit{informative}.

\begin{compactitem}
	\item \textbf{Semantic score}: This metric aims to evaluate the quality of the selected icons and themes.
	Since icons and themes are selected through keyword matching, the better the matching, the higher the semantic score is.
	For example, given the statement ``70\% students find math difficult,'' the icon of ``student'' should contribute to a higher semantic score than the icon of ``person''. Thus, we define the semantic score $\alpha_s$ as the average value of all the displayed icons and color palettes picked from the \textit{graphic} and \textit{color} modules using the Word2Vec~\cite{mikolov2013efficient} measure. Since we prefer infographics with graphic elements, we define $\alpha_s = 0$ if there are no matching icons or color palettes.
	\item \textbf{Visual score}: This metric aims to evaluate the visual appearance of an infographic.
	Since the aspect ratio of icons or the length of the description embedded in the infographic may not be ideal to the layout design, icons and text may be scaled to fit into the layout.
	Thus, this metric is designed to measure the empty space that these visual elements waste.
	The better they fit, the higher the visual score is.
	Thus, we define the visual score as:
	$$\alpha_v = \frac{\textrm{the area of all displayed elements}}{\textrm{the area of the canvas}}$$
	\item \textbf{Informative score}: Ideally, readers should easily recover the original information from the resulting infographic.
	However, due to the applicable constraints of these values, some information may be left out in infographics, which clearly should be avoided.
	For example, shorter descriptions may be used due to space limitations.
	Therefore, we propose the informative metric to evaluate the completeness of message delivery.
	Thus, we define the informative score as $\alpha_i=\sum_{\omega \in S}I(\omega)/|S|$, where $S$ denotes all the non-stop words in the input statement, and 
	$$I(\omega) = \begin{cases}
	1 & \textrm{ if $\omega$ appears in the result as an icon or word}\\
	0 & \textrm{ otherwise.}
	\end{cases}$$		
\end{compactitem}

The total score is defined as a weighted sum of all three scores: $\alpha = w_s\alpha_s+w_v\alpha_v+w_i\alpha_i$.
The default values of these weights are set to 0.25, 0.5, and 0.25, respectively, based on our experiments.

\subsubsection{User Refinement}

Although we realize that different users may have different tastes and aim to provide a variety of choices so that users can pick one that best meets their personal tastes, it is still possible that users need to perform minor touches to the final results.
Without compromising too much of the intuitiveness of our system, we decided to support users replacing elements in each infographic, including icons, color palettes, and descriptions. 
\new{By clicking on a recommended infographic, users can open a dialog, in which all the adopted icons, colors, and descriptions are listed. Then users can easily replace them with whatever alternatives they like. When users are satisfied with the result, they can easily save the template or infographic for the reuse purpose.}

\section{Evaluation}
In this section, we first present a set of examples to demonstrate the diverse designs that our system can automatically create.
To understand how general users perceive our system, we further conducted a set of casual interviews with a wide variety of audiences in two exhibit events.
The last assessment involved expert evaluation with three professional graphic designers.

\subsection{Sample Infographics}
To demonstrate the capabilities of Text-to-Viz, we present a variety of infographics created with our system (Figure~\ref{fig_examples}).
For example, Figure~\ref{fig_examples}(a)-(d) and Figure~\ref{fig_layout_example}(b) are all generated based on the same statement, ``More than 20\% of smartphone users are social network users.''
We can see that different templates can produce different infographics.
In particular, we can see that Figure~\ref{fig_layout_example}(b) is based on the layout blueprint illustrated in Figure~\ref{fig_layout_example}(a).
However, since the template does not reserve a place for the \textit{modifier} component, the generated infographic is less accurate than the others, and hence has a lower informative score.
Figure~\ref{fig_examples}(e) shows an example of a filled icon, while Figure~\ref{fig_examples}(f) shows an example of a tilted layout.
Figure~\ref{fig_examples}(g) and (h) demonstrate two examples of how our system handles proportion information in the form of ``\textit{m} in \textit{n}''.
Our system can either choose the correct number of icons to form a pictograph (Figure~\ref{fig_examples}(g)) or convert the information to a percentage number and show it with other visualizations (Figure~\ref{fig_examples}(h)).
Figure~\ref{fig_examples}(i) and (j) show that our color strategy can correctly select colors based on semantic information.
Since one of our color palettes has the descriptive keyword \textit{coffee}, this color palette will be ranked higher when choosing colors for infographics.
Figure~\ref{fig_examples}(k)-(o) demonstrate the results for showing multiple percentages.
Specifically, Figure~\ref{fig_examples}(k)-(m) show a \textit{comparison} case, in which proportion facts cannot be logically accumulated, while Figure~\ref{fig_examples}(n) and (o) show an \textit{accumulation} case.

\subsection{Casual User Interview in Exhibits}
\label{sec_casual}
Our system has been demonstrated to the public in two exhibits.
One was a half-day exhibit with an audience consisting mostly of journalists, and the other event was a two-day exhibit with an audience consisting of employees from all departments of a tech company, including sales, software developers, program managers, and public relation managers, etc.
In the two events, we received more than 80 different audience members in total. During each reception, we first introduced the background and system and gave them a short live demo to illustrate the idea. 
Then, we encouraged them to explore the system using their own statements. 
During the discussions, we focused on the following questions:
\begin{compactitem}
	\item Do you think this tool is useful? 
	\item Do you like the results generated?
	\item Do you have any suggestions considering your \old{typical work} background? 
\end{compactitem}

Overall, the feedback is overwhelmingly positive. 
When we explained the problem to the audience members, they understood immediately, since they had all encountered similar situations as described in Section~\ref{sec_intro}.
Since most of them did not have a design background, this tool would provide much greater convenience when they need infographics.
Several audience members described all the troubles they had gone through to build an infographic and how they had finally given up after several attempts, as they felt the tools were too complicated for them.
Given their different backgrounds, the audience members also suggested a variety of applications for this tool.
The journalists immediately saw the value of it for their reports.
Several public relation managers and project managers believed this tool would greatly help their presentations.
In particular, one development manager asked if we could support speech-to-infographics and said, ``It would be very interesting to see infographics popping up on screens when people are discussing in a meeting.''

One compromise we made for this system was creativity.
Since our infographic layouts are based on templates, we were worried that the output infographics may also lack creativity and feel homogeneous.
However, when we asked the audience how they felt about the output results, to our surprise, none of them unpromptedly noticed the homogeneity and creativity issue.
All of them commented that the outputs were very impressive and professional and that they would love to directly use them out of the box.
We suspected that this may be because they, again, lacked design experiences and only used our system momentary.
However, we did encounter some problems when we asked the users to try our system with their own statements.
The most critical problem was the recommended icons, which sometimes did not match the expectations from users.
For example, when a lab director input the statement: ``30\% secretaries wear glasses'', the recommended icon was a woman's skirt, which seemed inappropriate.
This was caused by the Word2Vec matching algorithm.
We also noticed that the audience members were the most sensitive to icons, which is probably because icons are more interpretable by non-designers, compared to other visual channels, such as \textit{layout}, and \textit{color}.
Therefore, we should put more effort into generating meaningful icons from text in the future, to ensure more satisfactory infographics overall.
Meanwhile, we allow users to manually replace icons as a temporary solution.

The most frequently asked question was when we would make this tool publicly available.
In addition, the audience also asked if we could support more types of statements.
Both requirements are expected since they can certainly help the audience members with their tasks.
On the other hand, we also collected some unexpected feedback, which indicated us a future direction for this project.
For example, one audience member asked if our algorithm could automatically adjust colors or styles to make it match an existing document or PowerPoint presentation.
Another data scientist suggested that we could perhaps generate infographics directly from a dataset.
By connecting to a data analysis module, our system can become more intelligent and then be used in many other scenarios.

\subsection{Expert Interview}
To understand the effectiveness of our system, we conducted a user study with three designers in a tech company. All of them had graduated (E1, E3) from or were enrolled (E2) in
professional schools of design disciplines. 
E1 majored in user experience design and had more than four years of design experience; E2 was a student majoring in industrial design with more than two years of design experience; and E3 majored in visual communications and had more than eight years of experience in design and two years of experience in photography. Their current roles are user interface designer and design intern.

The 60-minute study started with a 10-minute introduction of our tool, including the usages, several use cases, and the sample infographic designs it generated. After that, a semi-structured interview was conducted to understand the designers' views on the workflow, the tool's usability, and the resulting infographics.

Overall, the designers agreed that the tool has very promising usages. They felt that the results generated are very impressive and can be used for different purposes, such as posters, presentation slides, and data reports. All of them said they would like to use it when available.

Our participants appreciated the overall design of the workflow. They expressed the needs for using such a system in their daily work and felt that it would be very convenient to have text as input and a list of candidate designs as output. ``It can be used in many productivity tools to enhance data presentation,'' said one of the designers. Further, E1 suggested that it should be integrated into daily work settings,  which would improve the authoring process to be transparent and without interruption.  E1 also mentioned, ``If it is not well-integrated, I would like to have a copy-to-clipboard button to reduce the effort of using it.''
In terms of the authoring functions we have provided, such as changing icons and color themes, the experts felt that there is no need to incorporate more editing, saying, ``It is obvious that the target of this tool is not for producing very creative and elaborate designs, so it is great to keep it simple and clear.''

In terms of the quality of the generated infographics, the designers thought it was good enough considering its target user group.
E1 said, ``Users may be lazy to seek perfection. A relatively good design using little effort is sufficient in most cases.'' For most of the use cases, the designers felt that they were flexible and diverse enough. They even commented: ``The number of choices is \textit{`not the more the better'}. Too many choices may become a burden and make it hard to make decisions.'' They also gave advices on the resulting infographics. For example, E1 said, ``Some of the infographics involve more than two fonts, but I think using one is ideal.'' E3 thought it would be acceptable to sacrifice visualization accuracy for better visual appealingness. For example, for the number ``65\%'', E3 expressed that it would be better to fill in two out of three icons. They also made suggestions on further considering the semantic meaning of the icons. For example, when filling a cup or waterdrop icon with colors, it should be bottom-to-top instead of left-to-right.

\section{Discussion}
\subsection{Automatic Generation of Infographics}
Traditionally, infographic design is considered a highly creative task that can hardly be performed solely by computers.
Therefore, although many interactive tools have been developed to build infographics, they still maintain a manual editing process, in which designers lead and computers assist.
This is not a suitable paradigm for casual users, who are also a major category of infographic users overlooked by previous research.
Unlike designers, casual users neither possess design expertise nor are they familiar with advanced tools, which has also been repeatedly confirmed during our casual interviews (Section~\ref{sec_casual}).
Therefore, when they need an infographic in their presentations or reports, they must either go through the troubles of creating one by themselves, or ask a professional designer to make one for them.
Neither approach seems to have a good return-of-investment, especially when their expectations for the infographic are not high.
In our survey, we indeed found that, despite creative designs, there are simpler designs that are used repeatedly in various samples, which leads us to believe that many people (even designers) still favor simple and clear infographics, and do not like overly-embellished designs.
This scenario motivated our work: to generate sufficiently good infographics in a most natural way, using natural language statements, for casual users.

\old{In this paper, we propose a general pipeline and demonstrate it through a proof-of-concept system.
For the expression space, we collected samples that people have actually used in their presentations, identified the essential information, and built a CNN model to extract them automatically.
Although the statements may have different grammar structures, the internal extracted information is standardized and serves as a bridge between the expression space and visual space.
For the visual space, we propose a four-axes design space, namely, \textit{layout}, \textit{description}, \textit{graphic}, and \textit{color}.
Although these four dimensions were identified based on the infographics related to proportion facts, we believe that this decomposition of space is also suitable for other types of infographics.}

\old{Then naturally, the critical step would be finding good values along these four axes when given the extracted information.
The more appropriate the values found, the better the resulting infographics.
Although we have demonstrated a proof-of-concept system in this paper, the module implementations are still open for exploration and improvement in order to obtain better results.
In addition to individual module implementations, how this framework could be extended to support more types of information is also a challenging task.}

\subsection{Opportunities for New Usage}

The opportunities for this solution are enormous.
With little or no human interactions, infographics can now be produced effortlessly, hence attracting more and more casual users to take advantage of infographics in their daily life.
However, our proof-of-concept implementation may not be the best use scenario for this kind of technology, since it is still an interruptive experience.
Users have to stop their work at hand, open our system, enter the statement, then obtain a proper result to insert back into their work.
However, we anticipate that this tool can work seamlessly with productivity software suites, such as Office and iWork.
When users are working on their presentations and reports, the text analysis module can run silently in the background.
Once it detects a statement that can be augmented with infographics, a message would come prompt up to check with users whether they want to take a recommended infographic.
In this way, infographics can reach the broadest range of audience and be helpful to them.

Another interesting opportunity is poster authoring. 
An infographic poster often contains multiple infographics that are semantically related.
Although each infographic can be generated independently, combining them as a poster can deliver a stronger or more complex message.
To achieve this goal, additional considerations, such as typesetting and visual consistency, should also be put into this equation.

\new{In this paper, we demonstrate a proof-of-concept system that takes natural language statements as input.
However, we think it is possible to connect the system to other data sources, such as database or documents.
Although we may need to incorporate more machine learning techniques to extract interesting patterns first, it will also greatly expand the applicability of our approach, e.g., integrated into professional visual analytics systems for data analysis tasks.}

\subsection{Failure Cases and Limitations}
\new{During the development of our prototype, we also observed some failure cases, in which our system generates wrong or bad results. One critical source of these failure cases is the \textit{text analyzer}. For some complicated and long statements, such as ``funds for administration were limited to 15\% of the scholarship amount,'' the analyzer may not segment correctly and provide correct tags, which will eventually lead to incomprehensible infographics.
We believe more training data can mitigate this situation. 
In practical use, we can also increase the prediction threshold to reduce the number of inaccurate predictions to surface.
Another main source is our icon selector.
Many bad cases we observed are caused by inappropriately matched icons.
In addition, the length and wrapping of descriptions may also cause less pleasing or readable infographics.
For example, if a description is too long, our system will automatically select a small font size, which may result in an inharmonious or even illegible infographic.
Although we have integrated a ranking mechanism to help evaluate the overall quality of generated infographics, there are still many aspects of aesthetic to consider in the future.
}

\new{In addition, our design also has some limitations in terms of capability. 
The first and foremost limitation is that our current approach can only handle a relatively small set of information. 
However, there are various types of information that can be represented by infographics.
So far, our approach has to manually identify key information and visual representations type by type.	
Although in this paper, we demonstrate the auto-generation paradigm on one type of information (proportion facts), it is still unknown how this paradigm extends to other types of information.}
The second limitation is infographic expressiveness.
Clearly, this approach lacks human creativity and is based on a set of pre-designed infographic styles.
Although it is an open framework that allows users to add more materials to enrich the infographic designs, the resulting infographics are still limited.
The third limitation concerns expression ambiguity.
For example, consider these two statements: ``30\% of students are French; while 40\% are American,'' and ``30\% of students speak French, while 40\% speak English.''
We can see that in the first statement, the two percentages can be aggregated, which makes it is acceptable to only use one pie chart to visualize the information.
However, for the second statement, it is not appropriate to combine the two values in one pie chart, since there may be students who speak both French and English.
This kind of ambiguity is extremely hard to resolve using our current machine learning model.
Thus, a model that incorporates a deeper understanding of knowledge is required.
What adds to the problem is user intent.
Even for the former expression, users may still choose one or two pies to emphasize the aggregation or comparison, although both are reasonable visuals.
The user intent is impossible to infer from the above examples.
In our implementation, we do not tackle this ambiguity problem and leave the decision to users.

\old{The first limitation is infographic expressiveness.
Clearly, this approach lacks human creativity.
Although it is an open framework that allows users to add more materials to enrich the infographic candidates, the resulting infographics are still limited.
The second limitation is related to extendability. 
There are various types of information that can be represented by infographics.
So far, our approach has to manually identify key information and visual representations type by type.
Although in this paper, we demonstrate the auto-generation paradigm on one type of information (proportion facts), it is still unknown how this paradigm extends to other types of information.
The third limitation concerns expression ambiguity.
For example, consider these two statements: ``30\% of students are French; while 40\% are American,'' and ``30\% of students speak French, while 40\% speak English.''
We can see that in the first statement, the two percentages can be aggregated, which makes it is acceptable to only use one pie chart to visualize the information.
However, for the second statement, it is not appropriate to combine the two values in one pie chart, since there may be students who speak both French and English.
This kind of ambiguity is extremely hard to resolve using our current machine learning model.
Thus, a model that incorporates a deeper understanding of knowledge is required.
What adds to the problem is user intent.
Even for the former expression, users may still choose one or two pies to emphasize the aggregation or comparison, although both are reasonable visuals.
The user intent is impossible to infer from the above examples.
In our proof-of-concept implementation, we do not tackle this kind of ambiguity and leave the decision to users.}

\section{Conclusion and Future Work}

In this paper, we introduce a framework to automatically generate infographics and demonstrate its feasibility through a proof-of-concept system.
Our system takes a natural language statement on a proportion fact and translates it to a set of professional infographics with different styles and settings, which casual users select from or refine based on their personal preferences.
Our approach does not require a complex authoring process or design expertise.
The example results and user/designer interviews show the tool's usability and promise of being adopted in everyday life.

There are several avenues that are promising for future work.
As a proof-of-concept, our system works relatively well and is only for a specific type of information.
\new{
In the future, we would like to expand our work to support more types of statistical information or even other types of infographic, such as timelines and locations.}
\old{Clearly, we can expand this system to support more types of statistical information or even other types of infographics, such as timelines and locations.}
On the other hand, the implementation of our current system is also limited.
For example, many algorithms used here are still rule-based, such as icon selection, color selection, and ranking.
We believe it will be very interesting to explore more techniques (especially machine learning based techniques) to further improve the result quality of our current system.

\bibliographystyle{abbrv}
\bibliography{main}


%

\end{document}